%% file: twisted.tex
\documentclass[12pt,prd,showpacs, superscriptaddress,floatfix,nofootinbib]{revtex4}
\usepackage{graphics,epsf,amsmath,color}
\usepackage{axodraw}
\newcommand\cu{Physics Department, Columbia University, New York,
  NY 10027, USA}
\newcommand\soton{School of Physics and Astronomy, University of
  Southampton,\\ Southampton SO17 1BJ, UK}
\newcommand{\co}{{\cal O}}

\begin{document}
\bibliographystyle{apsrev}
\title{\boldmath{$K\to(\pi\pi)_{I=2}$} decays and twisted boundary conditions}
\author{C.h.\,Kim}\affiliation{\cu}
\author{C.T.\,Sachrajda}\affiliation{\soton}
\collaboration{RBC and UKQCD Collaborations}

\begin{abstract} We propose a new method to evaluate the Lellouch-L\"uscher factor which relates the $\Delta I=3/2$ $K\to\pi\pi$ matrix elements computed on a finite lattice to the physical (infinite-volume) decay amplitudes. The method relies on the use of partially twisted boundary conditions, which allow the s-wave $\pi\pi$ phase shift to be computed as an almost continuous function of the centre-of-mass relative momentum and hence for its derivative to be evaluated. We successfully demonstrate the feasibility of the technique in an exploratory computation.\end{abstract}


\vspace{-5.5in}\begin{flushright}
CU-TP-1194\\
\vspace{-0.1in}SHEP 1006
\end{flushright}

\maketitle


\section{Introduction}\label{sec:intro}
The precise computation of $K\to\pi\pi$ decay amplitudes in lattice simulations would be a very important step in our ability to evaluate non-perturbative strong interaction effects in hadronic processes. In particular it would allow us to  compute the quantity $\varepsilon'/\varepsilon$ which contains the QCD effects in direct CP-violating decays
and to understand the origin of the long-standing puzzle of the $\Delta I=1/2$ rule. Such calculations have not yet been achieved, primarily due to the technical difficulties in computing the $\Delta I=1/2$ $K\to\pi\pi$ matrix elements and the corresponding disconnected diagrams with sufficient precision.

In the centre-of-mass frame, the relation between $K\to\pi\pi$ matrix elements computed in a finite Euclidean volume and the physical decay amplitudes is given by the Lellouch-L\"uscher (LL) factor~\cite{Lellouch:2000pv,Lin:2001ek}. This relation was generalised to a moving frame, in which the total momentum of the two-pion system is non-zero, in~\cite{Kim:2005gf,Christ:2005gi}. The LL factor depends on $\delta^\prime(q^{\ast})$, the derivative of the s-wave phase-shift $\delta(q^\ast)$ with respect to the relative momentum in the centre-of-mass frame $q^\ast$. Whereas the phase-shift itself can be determined from the two-pion spectrum in the finite volume~\cite{Luscher:1986pf,Luscher:1991ux}, the derivative can only be determined by an interpolation or from estimates using chiral perturbation theory or models. The aim of this paper is to demonstrate that for $\Delta I=3/2$ transitions it is possible to calculate both the phase-shift and its derivative directly at the quark masses used in the lattice simulation. This can be achieved by using (partially) twisted boundary conditions.

This paper is organized as follows. In the next section, we summarise the theoretical background, recalling the expression for the LL-factor and describing the main properties of partially twisted boundary conditions. In this section we also present our procedure for the determination of the derivative of the two-pion phase shift. We then test our proposed procedure in a numerical simulation which is described in sec.\ref{sec:lattice} and the results are presented in sec.\ref{sec:anal}. Finally in sec.\ref{sec:con} we present our conclusions.

\section{Theoretical Framework}\label{sec:theory}

We start with a reminder of why it is necessary to compute the derivative of the phase shift.
In an arbitrary moving frame the $K\to\pi\pi$ matrix elements in finite and infinite volumes, which we denote by $M$ and $A$ respectively, are related by~\cite{Lellouch:2000pv,Lin:2001ek,Kim:2005gf,Christ:2005gi}
\begin{equation}\label{eq:ll}
|A|^2=8\pi V^2\frac{m_KE^2}{q^{\ast\,2}}\,\left\{\delta^\prime(q^\ast)+\phi^{P\,\prime}(q^\ast)\right\}\,|M|^2\,,
\end{equation}
where $E$ is the total energy, $E^2-P^2=4(m_\pi^2+q^{\ast\,2})$ where $\vec{P}$ is the total momentum, the $^\prime$ denotes differentiation with respect to $q^\ast$, $\delta$ is the s-wave $\pi\pi$ phase-shift and $V=L^3$ is the spatial volume. The explicit form of the kinematic function $\phi^P$ can be found in \cite{Kim:2005gf,Christ:2005gi} and we do not need to reproduce it here. Our aim in this paper is to explain how $\delta^\prime(q^\ast)$ can be calculated directly for $\Delta I=3/2$ transitions and to demonstrate the feasibility of the proposed procedure in an exploratory lattice computation. In this way the factor relating the finite and infinite volume matrix elements can be determined at the quark masses used in the simulation.

In order to evaluate the $I=2$ s-wave phase-shift and its derivative we compute the correlation function $\langle 0|\pi^-(t)\pi^-(t)\,\pi^+(0)\pi^+(0)|0\rangle$, where $\pi^+$ and $\pi^-$ are interpolating operators with the correct quantum numbers to create or destroy a $\pi^+$ meson. The time $t$ is chosen to be positive. The precise form of the interpolating operators used in our exploratory simulation will be described in sec.~\ref{sec:lattice} below; the general argument presented here does not depend on the choice of these operators. With periodic boundary conditions the total momentum of the two-pion system $\vec{P}$ can take the values $\vec{P}=(2\pi/L)\,\vec{n}$, where $\vec{n}$ is an vector of integers. From the energy levels of the two-pion system we can then deduce the corresponding phase shifts~\cite{Luscher:1986pf,Luscher:1990ux} using L\"uscher's quantisation formula:
\begin{equation}\label{eq:quantization}
\delta(q^\ast)+\phi^P(q^\ast)=n\pi\,,
\end{equation}
where $n$ is an integer. In typical simulations $2\pi/L$ is several hundred MeV and so the allowed momenta are quantized in large increments making it impossible to calculate reliably the derivative of the phase-shift.

Changing the boundary conditions changes the momentum spectrum and Bedaque proposed exploiting this fact to extend the range of momenta available in lattice applications~\cite{Bedaque:2004kc}~\footnote{See~\cite{Bedaque:2004kc} and~\cite{deDivitiis:2004kq} for references to earlier papers with related ideas.}. For example if we impose the following spatial boundary condition for the $u$-quark:
\begin{equation}
u(x_i+L)=e^{i\theta_i}\,u(x_i)\,,\qquad (i=1,2,3)
\end{equation}
then the momentum spectrum of a free $\pi^+$ meson is given by
\begin{equation}
p_i=n_i\,\frac{2\pi}{L}+\frac{\theta_i}{L}\,,
\end{equation}
where the $n_i$ are integers. For illustration let us imagine that only $\theta_1\neq 0$ and we now drop the suffix and denote $\theta_1$ by $\theta$ (the generalization to arbitrary $\vec{\theta}$ is conceptually straightforward). For $\theta<\pi$ the ground state of the two-$\pi^+$ system then has momentum $2\theta/L$, corresponding to each pion having momentum $\theta/L$. Boosting to the centre-of-mass frame we find that $q^\ast=0$ for all $\theta$ so that nothing has been gained by introducing the twisted boundary conditions. This is not true if we add some units of $2\pi/L$ to the momentum by performing the corresponding Fourier transform. In this case, $q^\ast$ does depend on $\theta$ so that by measuring the two-pion energies with different $\theta$ we obtain the phase-shifts at different values of $q^\ast$. Since $\theta$ can be incremented by arbitrarily small amounts, in this way we can obtain the derivative of the phase shift. This is the main point which we wish to make in this paper.

As an illustrative example let the $u$-quarks satisfy twisted boundary conditions with angle $\theta$, the $d$ quarks satisfy periodic boundary conditions and the two-pion state be given a further momentum of $-2\pi/L$ by performing the corresponding Fourier transform. The total momentum of the two-pion state is then $P_\theta=2(\theta-\pi)/L$. We imagine that the matrix elements of the operators appearing in the $\Delta I=3/2$ weak Hamiltonian have been measured with periodic boundary conditions (i.e. with $\theta=0$) with a total momentum of $-2\pi/L$ and that we want to evaluate the corresponding LL factor. We now measure the two-pion ground state energies $E_\theta$ for a range of values of $\theta$ and determine the corresponding centre-of-mass relative momentum $q^\ast$ in the standard way using $E_\theta^2-P_\theta^2=4(m_\pi^2+q^{\ast\,2}_\theta)$. The phase shift $\delta(q^\ast_\theta)$ is obtained from the generalization of the L\"uscher quantization condition to a moving frame (\ref{eq:quantization}) ~\cite{Rummukainen:1995vs,Kim:2005gf,Christ:2005gi} and its derivative is determined from the observed slope of the results obtained at different values of $\theta$ which can be chosen to be arbitrarily close together.

The method proposed in this paper is feasible with the use of \textit{partially twisted} boundary conditions~\cite{Sachrajda:2004mi,Bedaque:2004ax} in which the twisted boundary conditions are applied only to the valence quarks, whereas the sea quarks satisfy periodic boundary conditions. It was shown in \cite{Sachrajda:2004mi} that with partially twisted boundary conditions the finite-volume effects remain exponentially small. The practical advantage of not varying the boundary conditions for the sea quarks is that it is not necessary to generate a new ensemble of gauge configurations for every choice of twisting angle; indeed it is this feature which makes the method feasible. Partially twisted boundary conditions have been shown to satisfy the dispersion relation and used in the evaluation of the leptonic decay constant $f_\pi$~\cite{Flynn:2005in}, in the evaluation of the electromagnetic form factor of the pion \cite{Boyle:2007wg,Boyle:2008yd,Frezzotti:2008dr} and of $K\to\pi$ semileptonic decay amplitudes~\cite{Flynn:2008hd}.

It is convenient to implement the partially twisted boundary conditions, by introducing a change of quark field variables $\psi\to\tilde\psi$:
\begin{equation}
\psi(x) = e^{i{\vec{\theta}\cdot\vec{x} \over
L}}\tilde{\psi}(x)\quad\textrm{and}\quad
\bar{\psi}(x) = e^{-i{\vec{\theta}\cdot\vec{x} \over
L}}\bar{\tilde{\psi}}(x)\,,
\label{eq:change_of_variable}\end{equation}
so that the $\tilde{\psi}(x)$ satisfy periodic boundary conditions.
The hopping terms in the lattice fermion action now become (for $i=1,2,3$)
\begin{equation}
\overline{\tilde{\psi}}(x)
 \left[ e^{i{a\theta_i\over L}} U_i(x)(1-\gamma_i)\tilde{\psi}(x+\hat i)
      + e^{-i{a\theta_i\over L}}U^\dagger_i(x-\hat i)
           (1+\gamma_i)\tilde{\psi}(x-\hat i)
 \right].
\end{equation}
The quark propagator is obtained by inverting the Dirac operator
in the gauge field background \{$e^{i{a\theta_i\over L}}U_i(x)$\}
where the $\{U_i(x)\}$ are generated from a dynamical Markov chain with sea quarks.
The resulting Dirac operator is hermitian and standard conjugate gradient algorithms can be used to invert it.

We end this section with three simple observations.
\begin{enumerate}
\item [i)] With the prescription described above the phase shifts obtained at different values of $q^{\ast}$ are correlated which, as explained in section \ref{subsec:derivative}, allows for a more precise determination of the slope than would be the case if the errors on the individual points were independent of each other.
\item [ii)] The second observation is that
the choice of the $\pi^+\pi^+$ state is the most convenient one with which to obtain $I=2$ phase-shifts, in spite of the fact that kaons do not decay into $\pi^+\pi^+$. The $\pi^+\pi^+$ state automatically has $I=2$ whereas, because the twisting cancels in the $\pi^0$, it is not possible to symmetrize the $\pi^+\pi^0$ state to separate the $I=2$ and $I=1$ components. The Wigner-Eckart theorem allows us also to obtain $K^+\to\pi^+\pi^0$ matrix elements of the operators in the effective weak Hamiltonian by computing $K^+\to\pi^+\pi^+$ matrix elements of the corresponding $\Delta I_z=3/2$ operator:
\begin{equation}
\frac{1}{\sqrt{2}}\,\left\{\langle\pi^+(p_1)\pi^0(p_2)\,|+
\langle\pi^+(p_2)\pi^0(p_1)\,|\right\}
|O^{3/2}|\,K^+\rangle=
\frac32\langle \pi^+(p_1)\pi^+(p_2)\,|O^{\prime\,3/2}|\,K^+\rangle\,.
\end{equation}
$O^{3/2}$ is one of the three $\Delta I=3/2$ operators:
\begin{eqnarray}
O^{3/2}_{(27,1)}&=&(\bar{s}^id^i)_L\,\big\{(\bar{u}^ju^j)_L-(\bar{d}^jd^j)_L\big\}+(\bar{s}^iu^i)_L\,(\bar{u}^jd^j)_L \label{eq:o271def}\\
O^{3/2}_{7}&=&(\bar{s}^id^i)_L\,\big\{(\bar{u}^ju^j)_R-(\bar{d}^jd^j)_R\big\}+(\bar{s}^iu^i)_L\,(\bar{u}^jd^j)_R \label{eq:o732def}\\
O^{3/2}_{8}&=&(\bar{s}^id^j)_L\,\big\{(\bar{u}^ju^i)_R-(\bar{d}^jd^i)_R\big\}+(\bar{s}^iu^j)_L\,(\bar{u}^jd^i)_R\,,
\label{eq:o832def}
\end{eqnarray}
where $i$ and $j$ are colour indices, $(\bar{q}_1 q_2)_{L,R}=\bar{q}_1\gamma^\mu(1\mp\gamma^5)q_2$ and $O^{\prime\,3/2}$ is the corresponding operator with $\Delta I_z=3/2$:
\begin{equation}\label{eq:oprimedef}
O^{\prime\, 3/2}_{(27,1)}=(\bar{s}^id^i)_L\,(\bar{u}^jd^j)_L\,,\ O^{\prime\, 3/2}_{7}=(\bar{s}^id^i)_L\,(\bar{u}^jd^j)_R\,,\ O^{\prime\, 3/2}_{8}=(\bar{s}^id^j)_L\,(\bar{u}^jd^i)_R\,.
\end{equation}
\item[iii)] Finally we recall that a finite cubic box breaks rotational invariance and hence mixes different partial waves. For lattice $\pi^+\pi^+$ states with partially twisted boundary conditions in addition to the s-wave, there are also components with $l=2,4,\dots$ and the results presented here were obtained under the assumption that only scattering in the s-wave is important in the energy range of interest (which is a good approximation for $K\to\pi\pi$ decays).
\end{enumerate}

This concludes the demonstration that, in principle at least, it is possible to calculate the derivative of the s-wave $\pi\pi$ phase-shift, and hence the Lellouch-L\"uscher factor, at the masses and momenta used to calculate the finite-volume $K\to\pi\pi$ matrix element. In the following section we test this idea in an exploratory simulation.

\section{Details of the Lattice Computation}\label{sec:lattice}

We now report on our exploratory numerical study of the ideas proposed in the previous section. The computations are performed on a $16^3\times 32\times 8$ lattice with $N_f=2+1$ Domain Wall Fermions and the Iwasaki gauge action~\cite{Iwasaki:1985we} at $\beta=2.13$. Details of the simulation and properties of the ensembles are described in \cite{Antonio:2006px}. The bare strange quark mass is fixed at $am_s=0.04$ and we use two ensembles with light quark masses $am_u=am_d=0.04$ and 0.02. (We will denote the common light-quark mass by $m$ and the strange quark mass by $m_s$.) The lattice spacing was estimated in \cite{Antonio:2006px} from the static potential using the value $r_0=0.5$\,fm and was found to be $1/a=1.826(48)(94)$\,GeV (in the chiral limit). Since our goal is to demonstrate the feasibility of the method rather than to obtain precise physical results we will use this value whenever we want to convert our results to physical units. We find that the ``pion" masses are $am_\pi=0.364(2)$ for $am=0.02$ and 0.468(2) for $am=0.04$~\footnote{In order to simplify the notation, in the remainder of the discussion we use lattice units setting $a=1$.}.

There are 2595 thermalized trajectories generated for each ensemble\,\cite{Antonio:2006px}.
For this study when $m=0.04$ we use 250 configurations separated by 10 trajectories, whereas for $m=0.02$ we use
120 configurations separated by 20 trajectories. In order to reduce the statistical error, for each configuration we
measure correlation functions with two different sources (at $t=0$ and $t=16$). In addition for each choice of $\theta=|\vec\theta\,|$ we perform measurements on each configuration with $\vec\theta=(\theta,0,0),\,(0,\theta,0)$ and $(0,0,\theta)$. The autocorrelations are studied using the standard binning technique, in which the results from a number of neighbouring configurations are combined in bins and treated as independent measurements. The bin size is increased until the corresponding statistical error stabilizes; for our observables this is the case with bins of size 4 for $m=0.02$ and 7 for $m=0.004$ (recall that the configurations are separated by 20 and 10 trajectories respectively for the two masses). The bin sizes are a little larger than those found for the observables studied in \cite{Antonio:2006px}.


\subsection{Interpolating operators}

For the interpolating operator for each pion at the source we use the gauge-fixed wall source with momentum projection,
\begin{equation}\label{eq:pisource}
\co_\pi^{\rm i}(\vec{p}_u, \vec{p}_d) = \sum_{\vec{x}} \bar{u}(\vec{x}\,) A^\dagger(\vec{x}\,) e^{ i \vec{p}_u \cdot \vec{x}}  \gamma_5 \sum_{\vec{y}}   A(\vec{y}\,)   d (\vec{y}\,) e^{ i \vec{p}_d \cdot \vec{y}}\,,
\end{equation}
where $A(\vec{x})$ is the gauge fixing matrix. In this study
we use the Coulomb gauge. The superscript i in $\co_\pi^{\rm i}$ stands for \textit{initial}. At the sink we use a point source
\begin{equation}\label{eq:pisink}
\co_\pi^{\rm f}(\vec{p}\,)
 = \sum_{\vec{x}} \bar{u}(\vec{x}\,)  \gamma_5 d(\vec{x}\,) e^{ i \vec{p} \cdot \vec{x}}.
\end{equation}
The two-pion interpolating operators are then constructed as the product of the single pion ones,
\begin{equation}
\co_{\pi\pi}^{\rm i} = \co_\pi^{\rm i}(\vec{p}_u, \vec{p}_d) \co_\pi^{\rm i}(\vec{p}_u , \vec{p}_d+\vec{P}\,)
\quad\textrm{and}\quad
\co_{\pi\pi}^{\rm f} = \co_\pi^{\rm f}(\vec{p}\,) \co_\pi^{\rm f}(\vec{p}+\vec{P}\,)
\label{eq:two_pion_op}\end{equation}
where $\vec{P}=(2\pi/L)\,\vec{n}$ and $\vec{n}$ is a vector of integers. We choose both pions to be created (or annihilated) on the same time slice.


\subsection{Twisted boundary condition}

In this calculation partially twisted boundary conditions with twisting angle $\vec{\theta}$ are applied
to the $u$-quarks, while the $d$-quarks satisfy periodic boundary conditions.
Thus, the momenta of free $\pi^+$-mesons are quantized as
\begin{equation}\label{eq:momspectrum}
\vec{p}_{\pi^+} = \frac{\vec{\theta}}{L} + \frac{2\pi}{L}\vec{n}\ .
\end{equation}
We choose to introduce momenta only in a single direction, the $z$-direction say, and take a fixed $\vec{P}=(0,0,-2\pi/L)$ for a variety of values of $\theta$ ($\vec\theta\equiv(0,0,\theta)$) with $0<\theta<\pi/L$. The choice of $|\vec{P}|$ to be the smallest non-trivial Fourier momentum is motivated by the need to keep lattice artefacts as small as possible.

An advantage of the change of variables introduced in Eq.\,(\ref{eq:change_of_variable}) is that the phase factor for momentum $\theta/L$ is already absorbed in the quark field. Therefore, the same form of interpolating operator
can be used for all twisting angles.

 \subsection{The Diagrams}

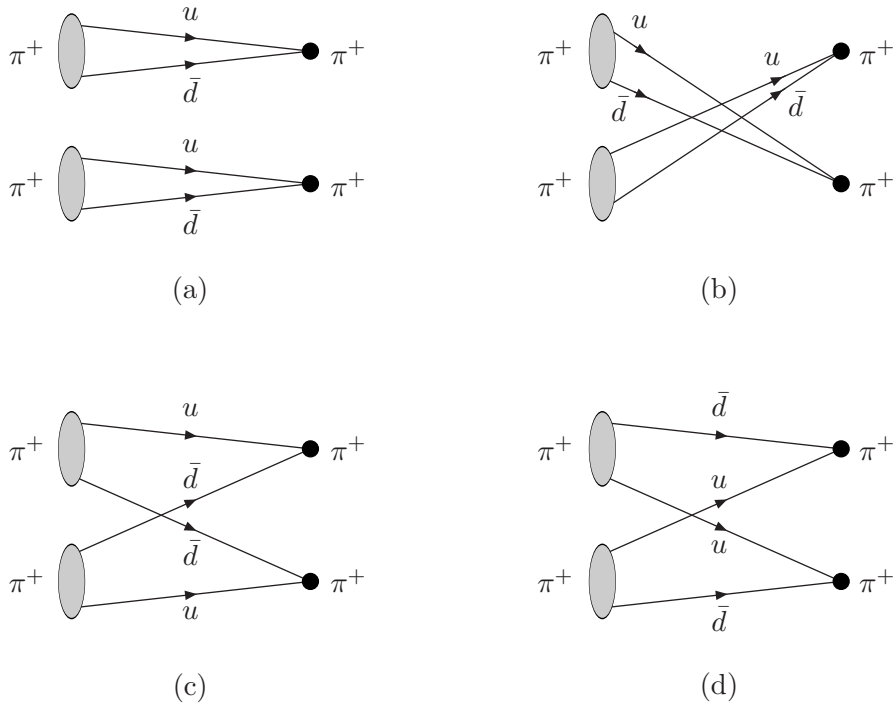
\begin{figure}[t]
\begin{center}
\begin{picture}(350,270)(0,-170)
\ArrowLine(10,85)(100,75)\ArrowLine(10,65)(100,75)
\ArrowLine(10,35)(100,25)\ArrowLine(10,15)(100,25)
\GCirc(100,75){3}{0}\GCirc(100,25){3}{0}
\GOval(10,75)(14,5)(0){0.8}\GOval(10,25)(14,5)(0){0.8}
\Text(55,87)[b]{\small{$u$}}\Text(55,65)[t]{\small{$\bar{d}$}}
\Text(55,37)[b]{\small{$u$}}\Text(55,15)[t]{\small{$\bar{d}$}}
\Text(55,-10)[t]{\small{(a)}}
\Text(0,75)[r]{\small$\pi^+$}\Text(0,25)[r]{\small$\pi^+$}
\Text(108,75)[l]{\small$\pi^+$}\Text(108,25)[l]{\small$\pi^+$}
\Line(210,85)(300,25)\Line(210,65)(300,25)
\Line(210,35)(300,75)\Line(210,15)(300,75)
\GCirc(300,75){3}{0}\GCirc(300,25){3}{0}
\GOval(210,75)(14,5)(0){0.8}\GOval(210,25)(14,5)(0){0.8}
\Text(225,84)[b]{\small{$u$}}\ArrowLine(222,77)(228,73)
\Text(217,58)[t]{\small{$\bar{d}$}}\ArrowLine(222,59.67)(228,57)
\ArrowLine(273,57)(279,61)\ArrowLine(273,63)(279,65.67)
\Text(275,70)[b]{\small{$u$}}\Text(284,60)[t]{\small{$\bar{d}$}}
\Text(255,-10)[t]{\small{(b)}}
\Text(200,75)[r]{\small$\pi^+$}\Text(200,25)[r]{\small$\pi^+$}
\Text(308,75)[l]{\small$\pi^+$}\Text(308,25)[l]{\small$\pi^+$}
\ArrowLine(10,-65)(100,-75)\ArrowLine(10,-85)(100,-125)
\ArrowLine(10,-115)(100,-75)\ArrowLine(10,-135)(100,-125)
\GCirc(100,-75){3}{0}\GCirc(100,-125){3}{0}
\GOval(10,-75)(14,5)(0){0.8}\GOval(10,-125)(14,5)(0){0.8}
\Text(55,-63)[b]{\small{$u$}}\Text(55,-90)[b]{\small{$\bar{d}$}}
\Text(55,-110)[t]{\small{$\bar{d}$}}\Text(55,-135)[t]{\small{$u$}}
\Text(55,-160)[t]{\small{(c)}}
\Text(0,-75)[r]{\small$\pi^+$}\Text(0,-125)[r]{\small$\pi^+$}
\Text(108,-75)[l]{\small$\pi^+$}\Text(108,-125)[l]{\small$\pi^+$}
\ArrowLine(210,-65)(300,-75)\ArrowLine(210,-85)(300,-125)
\ArrowLine(210,-115)(300,-75)\ArrowLine(210,-135)(300,-125)
\GCirc(300,-75){3}{0}\GCirc(300,-125){3}{0}
\GOval(210,-75)(14,5)(0){0.8}\GOval(210,-125)(14,5)(0){0.8}
\Text(255,-63)[b]{\small{$\bar{d}$}}\Text(255,-90)[b]{\small{$u$}}
\Text(255,-110)[t]{\small{$u$}}\Text(255,-135)[t]{\small{$\bar{d}$}}
\Text(255,-160)[t]{\small{(d)}}
\Text(200,-75)[r]{\small$\pi^+$}\Text(200,-125)[r]{\small$\pi^+$}
\Text(308,-75)[l]{\small$\pi^+$}\Text(308,-125)[l]{\small$\pi^+$}
\end{picture}
\caption{The four diagrams contributing to the $\pi^+\pi^+$ correlation function. The grey ovals represent the gauge-fixed wall sources for each of the $\pi^+$ interpolating operators and the black circles represent the local interpolating operators at the sink.\label{fig:pipi-cont}}
\end{center}\end{figure}

In the construction of the correlation function $\langle 0|\co_{\pi\pi}^{\dagger\,f}(t)\ \co_{\pi\pi}^i(0)\,|0\rangle$, where only the time variable is explicitly shown, there are four diagrams (or Wick contractions) as indicated in Fig.\ref{fig:pipi-cont}. The two \textit{disconnected} diagrams ($D$) in the first row of
Fig.\ref{fig:pipi-cont} each have two traces over the spinor and colour labels and for each gauge configuration are the product of two separate pion loops. The two \textit{connected} diagrams ($C$) in the second row of the figure have a single trace
over the spinor and colour labels and so the two-pion correlation function is the difference $D-C$, where the $-$ sign comes from Fermion statistics.


\section{Results}\label{sec:anal}

\begin{figure}[t]
\begin{center}
\epsfxsize=0.6\hsize
\epsfbox{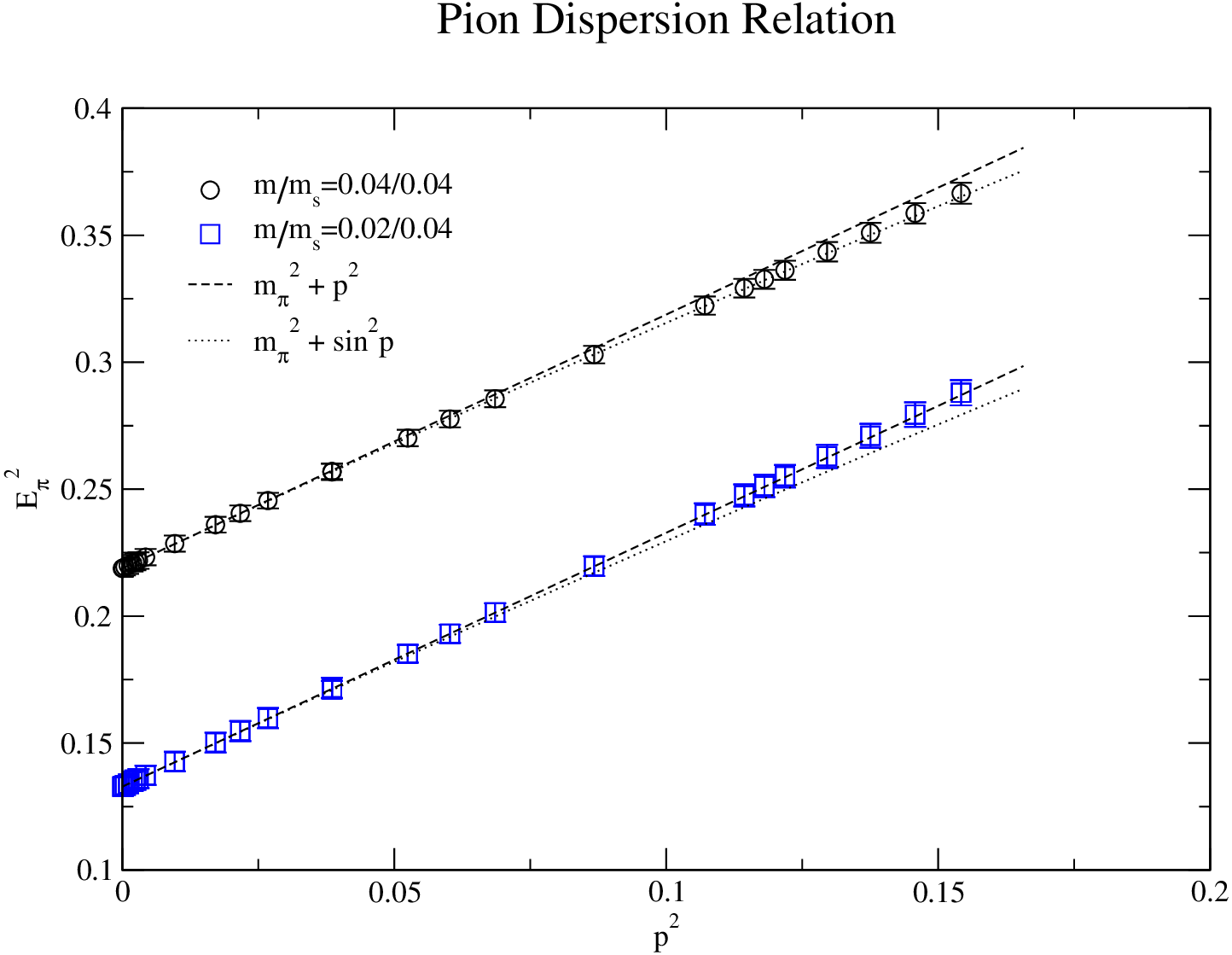}
\end{center}
\caption{Dispersion relation for pseudoscalar mesons with light quark masses $m_u=0.02$ and 0.04.}
\label{fig:dispersion}
\end{figure}

In this section we present the results of our numerical study. We measure the energies of the two-pion states for a range of values of $\theta$ and then use the L\"uscher quantization condition to determine the corresponding $\pi\pi$-phase shift. By choosing the twisting angles to be sufficiently close together we are then able to calculate the derivative of the phase shift and hence the corresponding LL factor. We start however, with a demonstration of the dispersion relation for a single pion as a function of the momentum induced by the twisting angle, $\vec{p}=\vec{\theta}/L$. In Fig.\ref{fig:dispersion} we plot the measured value of $E^2$ as a function of $p^2\equiv |\vec{p}\,|^2$ and compare it to the continuum $E^2=p^2+m^2$ dispersion relation as well as a possible discrete form, $E^2 = m^2 + \sin^2(p)$. The mass of the pion is determined in the standard way by fitting the correlation function at zero momentum so that the theoretical curves have no free parameters. The results are presented for two different values of the bare light quark mass $m=0.02$ and $0.04$ corresponding to pion masses $m_\pi=0.364(1)$ and $m_\pi=0.4676(8)$ respectively (to convert the pion msses into physical units, recall that the lattice spacing $a$ obtained from the Sommer scale is approximately given by $1/a=1.83(5)(9)$). In ref.\cite{Antonio:2006px} the residual mass for this action was found to be approximately $m_{\textrm{res}}\simeq 0.011$ and using this value the ratio of pion masses is as expected from the PCAC relation.

As is now well established (see for example ref.\,\cite{Flynn:2005in}), the dispersion relation is well satisfied. The small difference in the theoretical curves at larger values of the momentum is due to $O((ap)^4)$ terms. As the momenta approach $p^2=(2\pi/L)^2\simeq 0.154$, the smallest momentum which is accessible with periodic boundary conditions, there is a small discrepancy visible between the results of the lattice calculation and the continuum dispersion relation for the $m=0.04$ case. For the purposes of this feasibility study we neglect this difference.

\subsection{The energies of the two-pion states and the corresponding phase shifts.}
In this exploratory study we calculate the derivative of the phase-shift by using the simple ansatz
\begin{equation}\label{eq:derdef}
\frac{\partial \delta}{\partial q^\ast} \simeq \frac{\Delta \delta}{\Delta q^\ast}
=\frac{\delta(q^\ast_i) - \delta(q^\ast_{i-1})}{q^\ast_i - q^\ast_{i-1}},
\end{equation}
where $q^\ast_i$ and $q^\ast_{i-1}$ are neighbouring values of $q^\ast$. Since both the numerator and denominator are small quantities, it is necessary to exploit the fact that all the quantities on the right-hand side of (\ref{eq:derdef}) are calculated on the same configurations and are therefore correlated. This can be achieved by fitting all the necessary quantities, i.e. the $q^\ast_i$ and corresponding phase-shift, simultaneously.

\begin{figure}[t]
\begin{center}
\epsfxsize=0.6\hsize
\epsfbox{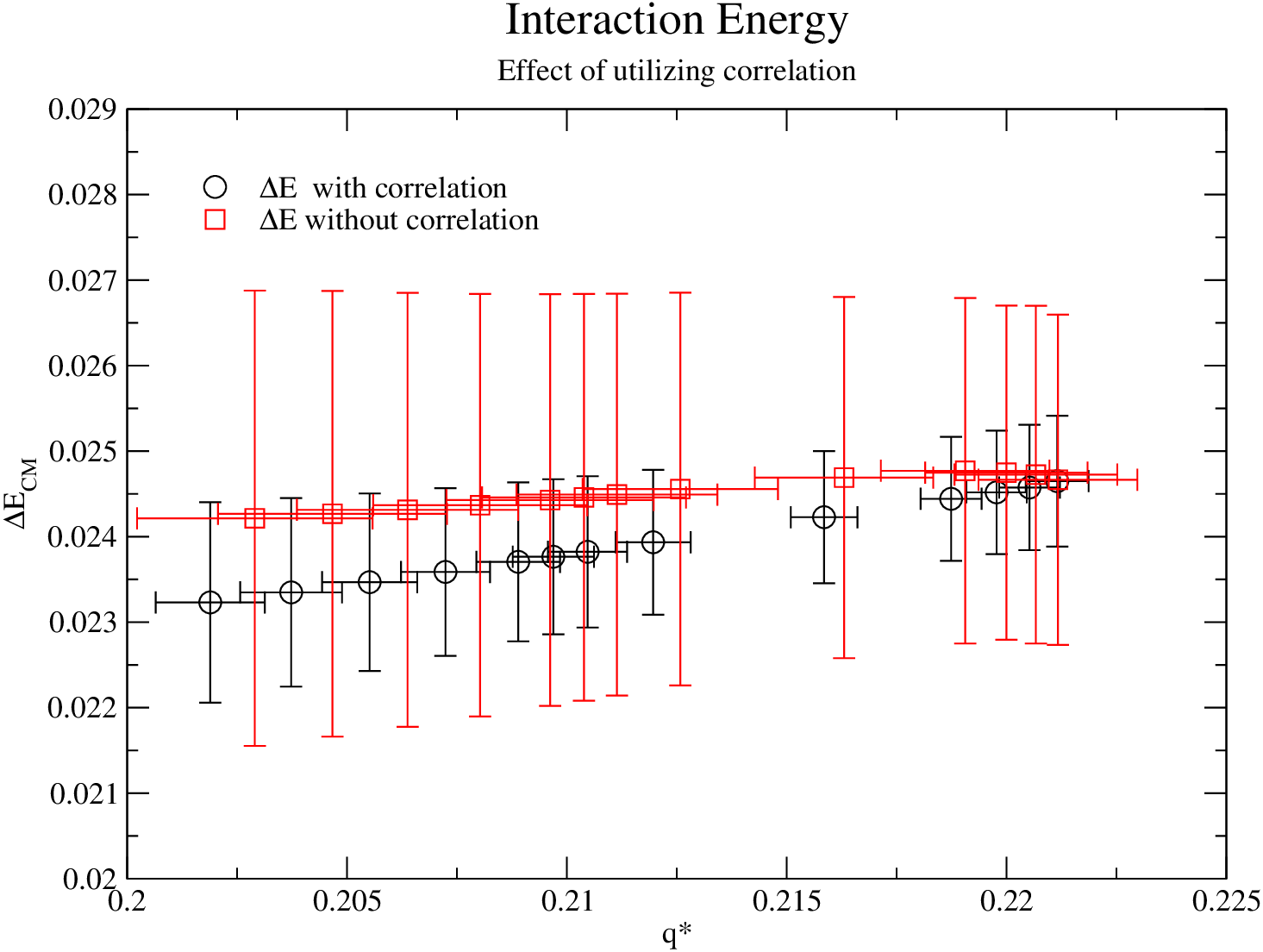}
\end{center}
\caption{The energy shift in the centre-of-mass frame as a function of $q^\ast$ obtained using eq.(\ref{eq:deltaE1}) (red squares) and (\ref{eq:deltaE2}) (black circles). The graph shows the advantage of exploiting the correlations in the two-pion and single-pion correlation functions.\label{fig:dE_corr_vs_uncorr}
}
\end{figure}

We start by considering the interaction energy
\begin{equation}
\Delta E = E_{\pi\pi} - \sqrt{m_\pi^2 + |\,\vec{p}_1|^2} - \sqrt{m_\pi^2 + |\,\vec{p}_2|^2}\,,
\label{eq:deltaE1}\end{equation}
where $E_{\pi\pi}$ is the energy of the two-pion state and $\vec{p}_1$ and $\vec{p}_2$ are the corresponding momenta of each of the pions if there were no interactions between them. The results for $\Delta E$ for $m=0.02$ obtained by measuring $E_{\pi\pi}$ and $m_\pi$ and boosting to the centre-of-mass frame are shown as the red squares in Fig.\ref{fig:dE_corr_vs_uncorr}.
From our numerical study however, we find that the errors in $\Delta E$ can be significantly reduced if instead of using eq.(\ref{eq:deltaE1}) we use
\begin{equation}
\Delta E = E_{\pi\pi} - E_{\pi,1} - E_{\pi,2}\,,
\label{eq:deltaE2}\end{equation}
where $E_{\pi,1}$ and $E_{\pi,2}$ are the measured energies of free pions with momenta $\vec{p}_1$ and $\vec{p}_2$ respectively. By performing a jackknife analysis of the right-hand-side of (\ref{eq:deltaE2}), we can take the correlations between $E_{\pi\pi}$ and $E_{\pi,1}+E_{\pi,2}$ into account. The results are plotted as the black circles in Fig.\ref{fig:dE_corr_vs_uncorr}, from which we see that the errors are decreased by more than a factor of 2. As the input two-pion energies into the L\"uscher quantisation formula we therefore take
\begin{equation}
E^{\rm in}_{\pi\pi} = \left\{E_{\pi\pi}-E_{\pi,1}-E_{\pi,2}\right\}
+\sqrt{m_\pi^2 + |\vec{p}_1|^2} + \sqrt{m_\pi^2 + |\vec{p}_2|^2}\,.
\end{equation}

The single-pion correlation functions are fit to the standard form
\begin{equation}
C_{\pi,i}(t) = B \cosh(E_{\pi,i}(t-T/2))\,,
\label{eq:cpiifit}\end{equation}
where $i=1$ or 2 and $T=32$ is the temporal extent of the lattice.

The leading behaviour of the two-pion correlation function takes the form:
\begin{equation}\label{eq:cpipi}
C_{\pi\pi}(t) = A\cosh(E_{\pi\pi} ( t - T/2 )) + C\cosh(D(t-T/2)),
\end{equation}
where the second term on the right-hand side is the contribution in which each of the two-pion interpolating operators destroys one pion and creates another, so that one pion propagates from $0$ to $t$ and the other from $t$ to $T$. Denoting the energies of the two pions in the second term by $E_{\pi,1}$ and $E_{\pi,2}$ then $D=|E_{\pi,1} -  E_{\pi,2}|$ and $C$ is proportional to $e^{-(  E_{\pi,1} +  E_{\pi,2} )T/2}$. In the first term $A$ is proportional to $e^{-E_{\pi\pi} T/2}$
and so, except in the vicinity of $t=T/2$, the first term in (\ref{eq:cpipi}) dominates the second.

By simultaneously fitting $C_{\pi,1}(t)$, $C_{\pi,2}(t)$ and $C_{\pi\pi}(t)$ to the forms in eqs.\,(\ref{eq:cpiifit}) and (\ref{eq:cpipi}) all the necessary quantities can be determined. We find however that the errors are reduced if we simultaneously fit the correlation functions $C_{\pi,1}(t)$, $C_{\pi,2}(t)$ and the ratio $R_{\pi\pi}(t)\equiv C_{\pi\pi}(t)/C_{\pi,1}(t)C_{\pi,2}(t)$ and the results quoted in the tables were obtained in this way. (To speed up the minimization of $\chi^2$, we found it useful to first fit $C_{\pi,1}(t)$, $C_{\pi,2}(t)$ and $C_{\pi\pi}(t)$ and then to use the resulting parameters as the starting values in the fits for $C_{\pi,1}(t)$, $C_{\pi,2}(t)$ and $R_{\pi\pi}(t)$.)

The fit ranges for $C_{\pi,1}(t)$ and $C_{\pi,2}(t)$
have already been determined from the standard effective mass plots. For the two-pion correlation functions $C_{\pi\pi}(t)$ we find that the range $t\in(8,17)$ is consistent for all twisting angles and for both values of the quark mass and we use the same range for $R_{\pi\pi}(t)$. Since our final goal is to determine $\partial \delta / \partial q^\ast$, having a single fitting range for all the twisting angles simplifies the analysis.

\begin{table}
\begin{center}
\input{two_pi_table002_1.tex}
\caption{Results for the total two-pion energy $E_{\pi\pi}$, the energy shift $\Delta E$, the relative centre-of-mass momentum $q^\ast$ and the $I=2$ s-wave phase shift $\delta_s^{I=2}$ for quark masses $m/m_s=0.02/0.04$. The total momentum $P=2(\pi-\theta)/L$.}
\label{tab:E002}
\end{center}
\end{table}

\begin{table}
\begin{center}
\input{two_pi_table004_1.tex}
\caption{Results for the total two-pion energy $E_{\pi\pi}$, the energy shift $\Delta E$, the relative centre-of-mass momentum $q^\ast$ and the $I=2$ s-wave phase shift $\delta_s^{I=2}$ for quark masses $m/m_s=0.04/0.04$. The total momentum $P=2(\pi-\theta)/L$.}
\label{tab:E004}
\end{center}
\end{table}

\begin{figure}[t]
\begin{center}
\epsfxsize=0.6\hsize
\epsfbox{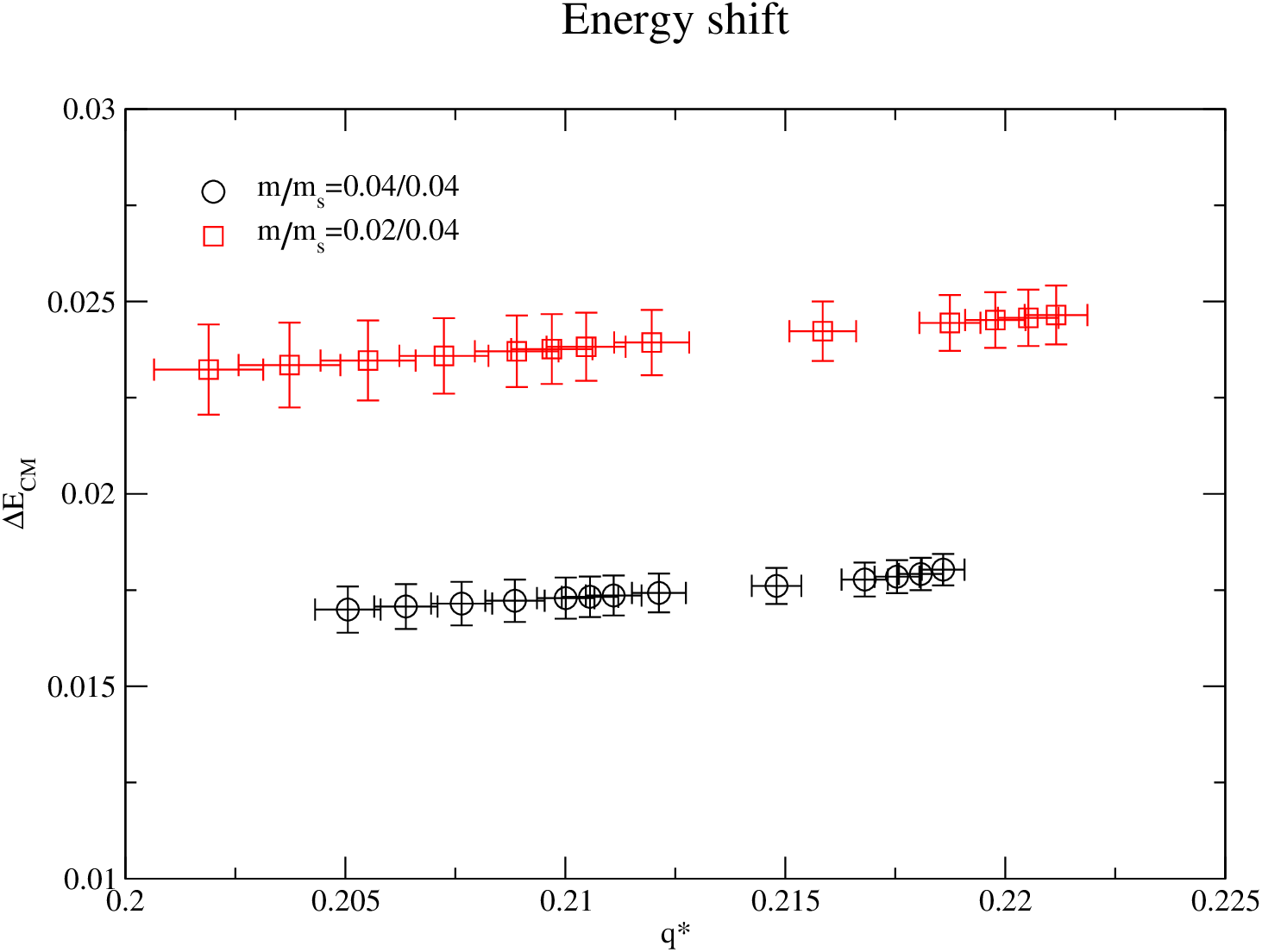}
\end{center}
\caption{The energy shift in the centre-of-mass frame as a function of $q^\ast$ at the two values of the light-quark masses.}
\label{fig:delta_E}
\end{figure}

\begin{figure}[t]
\begin{center}
\epsfxsize=0.6\hsize
\epsfbox{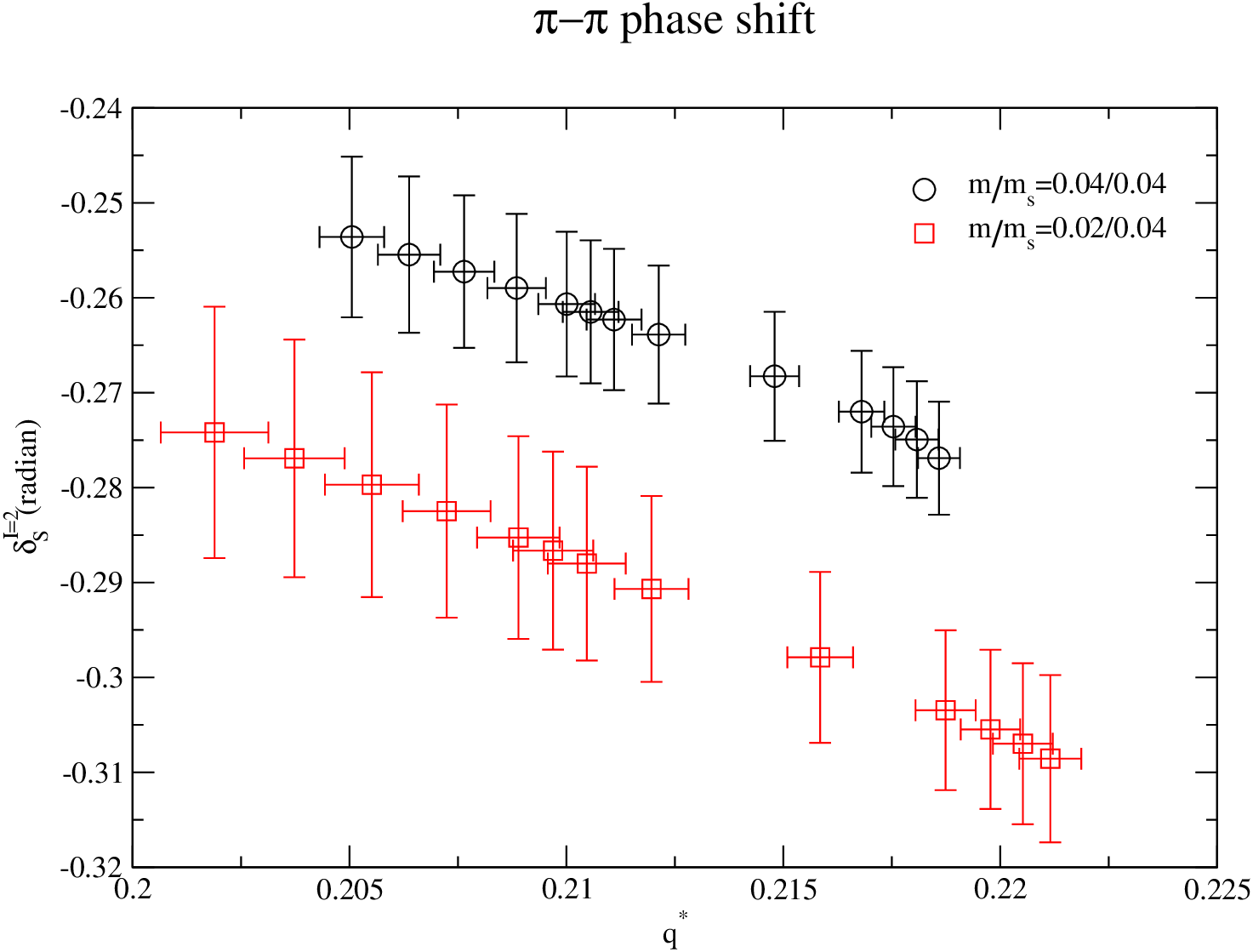}
\end{center}
\caption{The $I\!=2\ \pi\pi$ phase shift deduced from the measured energy shifts.}
\label{fig:phase}
\end{figure}

The energies of the $\pi\pi$ states are given in Tables.\ref{tab:E002} and \ref{tab:E004} for the two values of the light quark mass. In these tables we also give the values of the energy shift $\Delta E$,
the relative momentum in the centre-of-mass frame $q^\ast$ and the phase shift. The state with the largest total momentum has the largest total energy as expected, but it also has the smallest relative momentum $q^\ast$ because of the boost factor. In Fig.\ref{fig:delta_E} and Fig.\ref{fig:phase},
we plot $\Delta E$ in the centre-of-mass frame and the phase-shift as a function of $q^\ast$. $\Delta E_{\textrm{CM}}$ is obtained from the measured value of $\Delta E$ in the moving frame (see eq.(\ref{eq:deltaE2})\,) by $\Delta E_{\textrm{CM}}=\gamma\,\Delta E$, where the boost factor $\gamma=1/\sqrt{1-v^2}$ and $v=|\vec{p}_1+\vec{p_2}|/E_{\pi\pi}$.
As expected, the energy shift $E_{\textrm{CM}}$ is positive and increases as the relative momentum $q^\ast$ increases. The phase shift is negative reflecting the repulsive nature of the $I=2$ two-pion interaction.

%

\subsection{The derivative of the phase shift and the Lellouch-L\"uscher Factor}\label{subsec:derivative}
Once the phase shift has been determined as a function of $q^\ast$, the derivative can be calculated
and the Lellouch-L\"uscher factor can be estimated. The values of the phase shift at different $q^\ast$ are correlated of course and, as we shall see, this results in a smaller error on the slope than might have been expected simply by looking at the error bars on a pair of neighbouring points and considering them to be independent. This is because the results for the phase-shifts at different $q^\ast$ move up and down together as we sample different configurations, leading to a smaller variation of the slope.

There are many ways in which the derivative can be determined numerically. We choose to use the most simple-minded one, i.e. the simple ansatz in eq.(\ref{eq:derdef}).
Note that the ansatz provides the derivative most accurately at the mid-point between $q_i$ and $q_{i-1}$. We therefore evaluate $\phi^{P\,\prime}(q^\ast)$ in the Lellouch-L\"uscher factor (\ref{eq:ll}) at the same point, $(q_i+q_{i-1})/2$.

\begin{table}
\begin{center}
\input{two_pi_table002_2.tex}
\caption{Results for the derivative of the phase shift, the derivative of the kinematic function $\phi^P$ and the Lellouch-L\"uscher factor for quark masses $m/m_s=0.02/0.04$. The Lellouch-L\"uscher factor is presented here without the overall $V^2$ factor (see eq.\,(\ref{eq:ll})).}
\label{tab:mult_factor_002}
\end{center}
\end{table}

\begin{table}
\begin{center}
\input{two_pi_table004_2.tex}
\caption{Results for the derivative of the phase shift, the derivative of the kinematic function $\phi^P$ and the Lellouch-L\"uscher factor for quark masses $m/m_s=0.04/0.04$. The Lellouch-L\"uscher factor is presented here without the overall $V^2$ factor (see eq.\,(\ref{eq:ll})).}
\label{tab:mult_factor_004}
\end{center}
\end{table}
\begin{figure}[t]
\begin{center}
\epsfxsize=0.75\hsize
\epsfbox{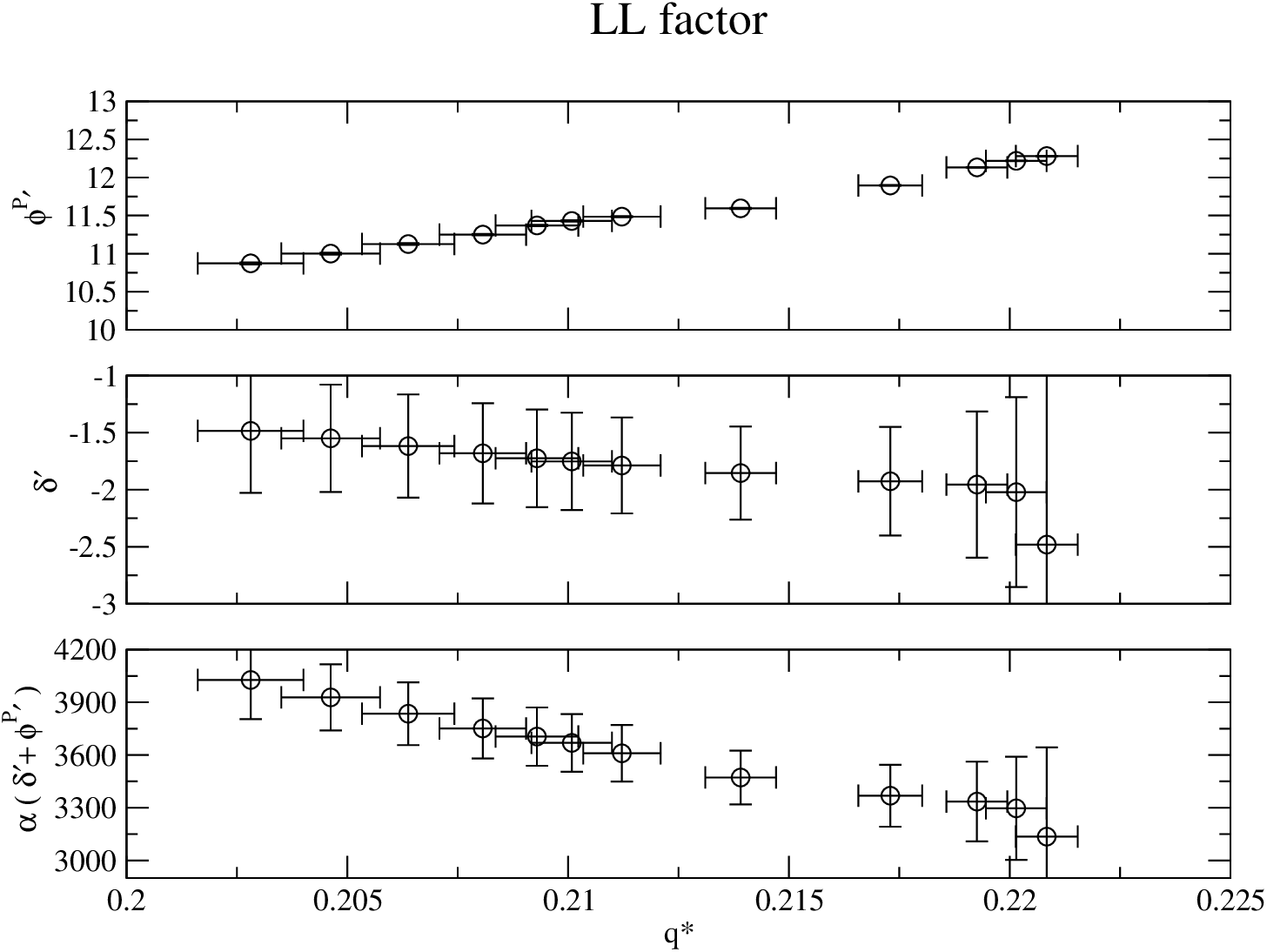}
\end{center}
\caption{The derivative of the phase-shift, $\phi^{P\,\prime}$ and the  Lellouch-L\"uscher factor for quark masses $m=0.02$ and $m_s=0.04$. $\alpha=8\pi m_KE^2/q^{\ast\,2}$ (see eq.(\ref{eq:ll}) and, as elsewhere in this paper, we do not include the overall factor of $V^2$ in our numerical results).}
\label{fig:LL_f_002}
\end{figure}

\begin{figure}[t]
\begin{center}
\epsfxsize=0.75\hsize
\epsfbox{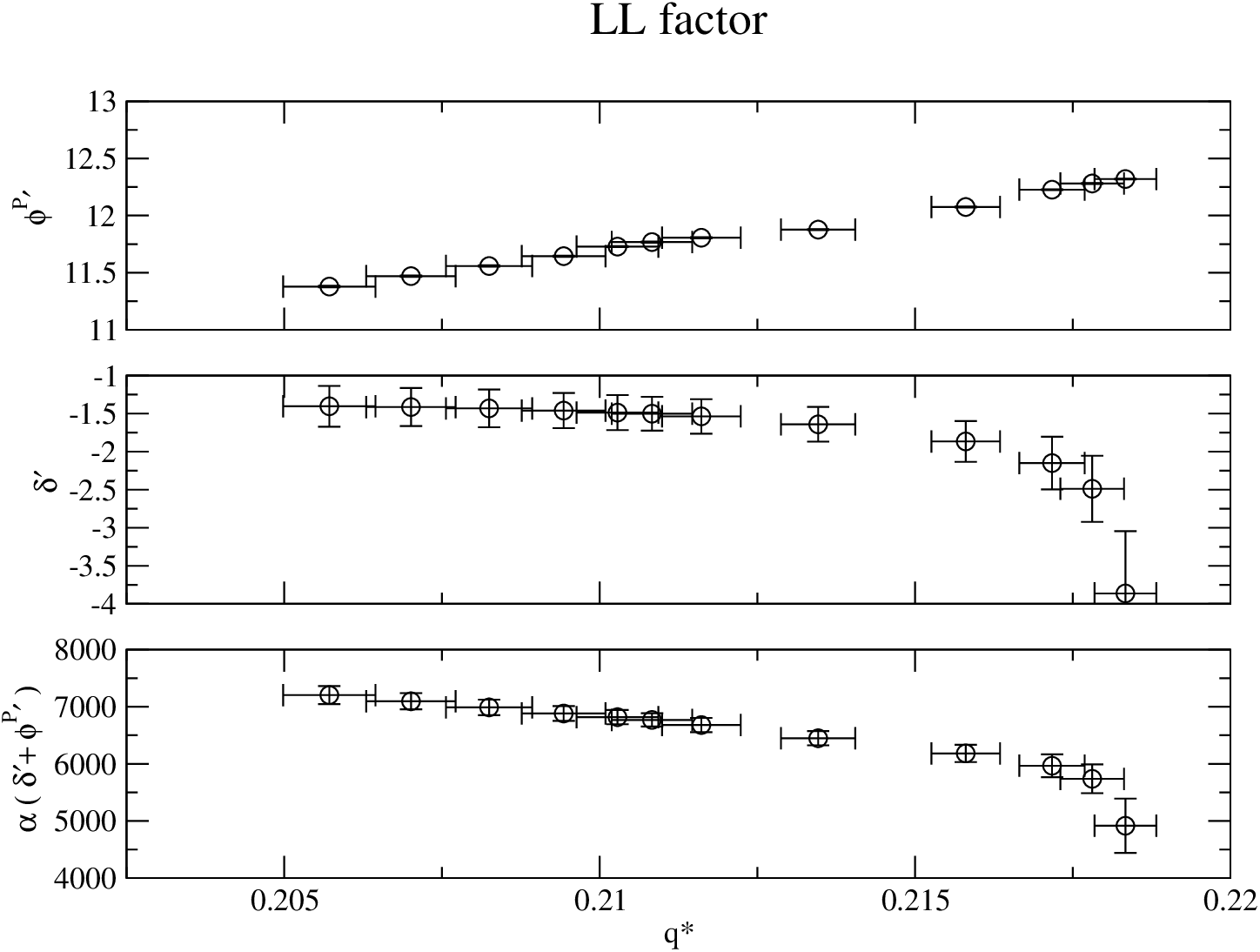}
\end{center}
\caption{The derivative of the phase-shift, $\phi^{P\,\prime}$ and the  Lellouch-L\"uscher factor for quark masses $m=m_s=0.04$. $\alpha=8\pi V^2m_KE^2/q^{\ast\,2}$ (see eq.(\ref{eq:ll}) and, as elsewhere in this paper, we do not include the overall factor of $V^2$ in our numerical results).}
\label{fig:LL_f_004}
\end{figure}

The results for the derivative of the phase-shift, together with the derivative of the kinematical function $\phi^P$ which is also an ingredient of the LL factor (see (\ref{eq:ll})) and the LL factor are presented in Tables \ref{tab:mult_factor_002} and \ref{tab:mult_factor_004}
and plotted in Figures \ref{fig:LL_f_002} and \ref{fig:LL_f_004}. Our final results for the Lellouch-L\"uscher factor have errors in the range 5-10\% when the light quark mass is 0.04 and 10-20\% when it is 0.02.
The final error in the LL factor comes predominantly from that in $\partial \delta /\partial q^*$, which contains the finite-volume corrections due to the interactions of the two pions. We see from Tables \ref{tab:mult_factor_002} and \ref{tab:mult_factor_004} that although $\partial \phi^P/ \partial q^*$ is larger than $\partial \delta /\partial q^*$ they have the opposite sign, so that the effect of the latter is indeed significant.

\section{Conclusions}\label{sec:con}

In this paper we have proposed a method for calculating the derivative of the s-wave $I=2$ $\pi\pi$ phase shift directly in lattice calculations. This enables us to evaluate the Lellouch-L\"uscher factor which relates the finite-volume $K\to\pi\pi$ matrix elements to the corresponding physical decay amplitudes at the quark masses and lattice volumes being simulated. The method relies on the use of partially twisted boundary conditions, allowing for the two-pion energies, and hence the corresponding phase-shifts, to be evaluated as a function of the centre-of-mass relative momenta.

The feasibility of the method was successfully tested in an exploratory computation. The phase shift was calculated as an almost continuous function of $q^\ast$ and the correlations between the points in Fig.\,\ref{fig:phase} allows a reliable determination of the LL factor as illustrated in figs.\,\ref{fig:LL_f_002} and \ref{fig:LL_f_004}. Although the volume of our lattice was small ($L\simeq 1.9$\,fm), this study does give us confidence that the technique can now be applied to a more physically realistic computation of $\Delta I=3/2$ $K\to\pi\pi$ matrix elements, such as that currently being undertaken by the RBC-UKQCD collaboration\,\cite{Christ:2009ev}.

We see from Tables \ref{tab:mult_factor_002} and \ref{tab:mult_factor_004} that although $\theta$ varies from 0 to $\pi$ and the total momentum varies from $2\pi/L$ to 0, the boost back to the centre-of-mass frame results in a very limited range of values of $q^\ast$ and hence we only evaluate the phase-shift and its derivative in this limited range. Nevertheless, this method allows us to evaluate the Lellouch-L\"uscher factor for $\Delta I=3/2$ $K\to\pi\pi$ decays for any kinematics which can be used with periodic or antiperiodic boundary conditions. For example, one could imagine calculating the $K\to\pi\pi$ amplitude with total momentum $2\pi/L$ and then use the technique proposed in this paper to determine the corresponding Lellouch-L\"uscher factor. In this case, the derivative would be evaluated from the behaviour of the phase-shift with small twisting angles. As a second example imagine introducing anti-periodic boundary conditions for the $u$ quark in the $z$-direction say, and calculating the matrix element between a kaon at rest and two pions with momenta $\pm \pi/L$. In this case one would determine the derivative from the behaviour of the phase-shift with twisting angle $\theta_z$ close to $\pi$.

Disappointingly, as explained in \cite{Sachrajda:2004mi}, the breaking of isospin symmetry induced by choosing different boundary conditions for the $u$ and $d$ quarks means that the method cannot be applied to $\Delta I=1/2$ transitions. In that case, even in the free theory, the $\pi^+\pi^-$ state has a different energy from the $\pi^0\pi^0$ state.

\subsection*{Acknowledgements} We warmly thank our colleagues from the RBC and UKQCD collaborations for their interest and encouragement. C.T.S. acknowledges support from STFC Grant ST/G000557/1 and EU contract MRTN-CT-2006-
035482 (Flavianet).

\bibliography{twisted}

\end{document}

%% file: two_pi_table002_1.tex
\begin{tabular}{cccccc}
 $\theta$ & $PL$ & $E_{\pi\pi}$ & $\Delta E$ & $q^*$ & $\delta^{I=2}_S$ \\ \hline
0 &  $ 2 \pi$  &   0.9220(36) & 0.0210(10) & 0.2019(12) & -0.274(13) \\ \hline
$\frac{ \pi}{18} $   &$\frac{17 \pi}{9} $   &   0.9145(36) & 0.0213(10) & 0.2037(11) & -0.277(12) \\ \hline
$\frac{ \pi}{9} $   &$\frac{16 \pi}{9} $   &   0.9075(35) & 0.02161(95) & 0.2055(10) & -0.280(11) \\ \hline
$\frac{ \pi}{6} $   &$\frac{5 \pi}{3} $   &   0.9008(35) & 0.02193(91) & 0.2072(10) & -0.282(11) \\ \hline
$\frac{2 \pi}{9} $   &$\frac{14 \pi}{9} $   &   0.8946(34) & 0.02224(87) & 0.20889(95) & -0.285(10) \\ \hline
$\frac{ \pi}{4} $   &$\frac{3 \pi}{2} $   &   0.8917(34) & 0.02239(85) & 0.20969(92) & -0.287(10) \\ \hline
$\frac{5 \pi}{18} $   &$\frac{13 \pi}{9} $   &   0.8888(34) & 0.02254(83) & 0.21047(89) & -0.288(10) \\ \hline
$\frac{ \pi}{3} $   &$\frac{4 \pi}{3} $   &   0.8835(34) & 0.02283(80) & 0.21196(85) & -0.2907(97) \\ \hline
$\frac{ \pi}{2} $   & $  \pi$  &   0.8700(35) & 0.02358(75) & 0.21585(75) & -0.2979(89) \\ \hline
$\frac{2 \pi}{3} $   &$\frac{2 \pi}{3} $   &   0.8604(33) & 0.02415(71) & 0.21874(69) & -0.3035(84) \\ \hline
$\frac{3 \pi}{4} $   &$\frac{ \pi}{2} $   &   0.8570(32) & 0.02435(71) & 0.21977(68) & -0.3055(83) \\ \hline
$\frac{5 \pi}{6} $   &$\frac{ \pi}{3} $   &   0.8546(29) & 0.02450(73) & 0.22052(68) & -0.3070(84) \\ \hline
 $  \pi$  &0 &    0.8526(29) & 0.02465(76) & 0.22115(71) & -0.3086(88) \\ \hline
\end{tabular}

%% file: two_pi_table004_1.tex
\begin{tabular}{cccccc}
 $\theta$ & $PL$ & $E_{\pi\pi}$ & $\Delta E$ & $q^*$ & $\delta^{I=2}_S$ \\ \hline
0 &  $ 2 \pi$  &   1.0891(28) & 0.01584(56) & 0.20505(74) & -0.2536(84) \\ \hline
$\frac{ \pi}{18} $   &$\frac{17 \pi}{9} $   &   1.0829(29) & 0.01603(54) & 0.20638(71) & -0.2554(82) \\ \hline
$\frac{ \pi}{9} $   &$\frac{16 \pi}{9} $   &   1.0771(29) & 0.01622(53) & 0.20764(69) & -0.2572(80) \\ \hline
$\frac{ \pi}{6} $   &$\frac{5 \pi}{3} $   &   1.0716(30) & 0.01639(52) & 0.20885(67) & -0.2590(78) \\ \hline
$\frac{2 \pi}{9} $   &$\frac{14 \pi}{9} $   &   1.0664(30) & 0.01656(51) & 0.21001(65) & -0.2607(76) \\ \hline
$\frac{ \pi}{4} $   &$\frac{3 \pi}{2} $   &   1.0639(29) & 0.01664(50) & 0.21056(64) & -0.2615(75) \\ \hline
$\frac{5 \pi}{18} $   &$\frac{13 \pi}{9} $   &   1.0616(30) & 0.01672(50) & 0.21110(63) & -0.2623(74) \\ \hline
$\frac{ \pi}{3} $   &$\frac{4 \pi}{3} $   &   1.0571(30) & 0.01688(48) & 0.21213(61) & -0.2639(72) \\ \hline
$\frac{ \pi}{2} $   & $  \pi$  &   1.0459(30) & 0.01729(46) & 0.21480(56) & -0.2683(67) \\ \hline
$\frac{2 \pi}{3} $   &$\frac{2 \pi}{3} $   &   1.0379(30) & 0.01763(43) & 0.21681(52) & -0.2720(64) \\ \hline
$\frac{3 \pi}{4} $   &$\frac{ \pi}{2} $   &   1.0351(29) & 0.01777(42) & 0.21754(51) & -0.2736(62) \\ \hline
$\frac{5 \pi}{6} $   &$\frac{ \pi}{3} $   &   1.0332(29) & 0.01788(41) & 0.21808(49) & -0.2749(61) \\ \hline
 $  \pi$  &0 &    1.0318(29) & 0.01803(40) & 0.21859(48) & -0.2769(59) \\ \hline
\end{tabular}

%% file: two_pi_table002_2.tex
\begin{tabular}{cccc}
$q^*$ & $\partial\delta / \partial q^*$ &  $\partial\phi^P / \partial q^*$ & LL factor \\ \hline
0.2028(11) & -1.49(54) & 10.873(13) & 4027(224) \\ \hline
0.2046(11) & -1.55(46) & 11.000(12) & 3928(189) \\ \hline
0.2064(10) & -1.62(45) & 11.126(11) & 3835(179) \\ \hline
0.20807(98) & -1.68(43) & 11.250(10) & 3751(172) \\ \hline
0.20929(93) & -1.73(42) & 11.3707(92) & 3704(166) \\ \hline
0.21008(91) & -1.75(42) & 11.4297(87) & 3669(165) \\ \hline
0.21122(87) & -1.79(42) & 11.4867(82) & 3609(162) \\ \hline
0.21391(80) & -1.85(40) & 11.5953(73) & 3472(153) \\ \hline
0.21730(72) & -1.93(47) & 11.8962(54) & 3368(176) \\ \hline
0.21926(68) & -1.96(63) & 12.1328(46) & 3335(227) \\ \hline
0.22015(68) & -2.02(83) & 12.2188(45) & 3296(294) \\ \hline
0.22084(70) & -2.4(1.5) & 12.2813(47) & 3135(508) \\ \hline
\end{tabular}

%% file: two_pi_table004_2.tex
\begin{tabular}{cccc}
$q^*$ & $\partial\delta / \partial q^*$ &  $\partial\phi^P / \partial q^*$ & LL factor \\ \hline
0.20572(73) & -1.40(26) & 11.3786(67) & 7204(159) \\ \hline
0.20701(70) & -1.41(25) & 11.4698(63) & 7095(143) \\ \hline
0.20825(68) & -1.43(24) & 11.5585(58) & 6988(137) \\ \hline
0.20943(66) & -1.46(22) & 11.6444(55) & 6882(129) \\ \hline
0.21028(64) & -1.49(23) & 11.7284(51) & 6817(126) \\ \hline
0.21083(63) & -1.50(22) & 11.7684(50) & 6769(116) \\ \hline
0.21161(62) & -1.54(22) & 11.8062(48) & 6679(125) \\ \hline
0.21346(58) & -1.64(22) & 11.8767(43) & 6448(125) \\ \hline
0.21580(54) & -1.87(26) & 12.0750(37) & 6181(152) \\ \hline
0.21717(51) & -2.15(34) & 12.2271(35) & 5965(200) \\ \hline
0.21781(50) & -2.49(43) & 12.2810(35) & 5737(252) \\ \hline
0.21833(49) & -3.86(82) & 12.3196(35) & 4915(476) \\ \hline
\end{tabular}